\documentclass[sigconf,screen]{acmart}

\setcopyright{cc}
\setcctype{by}
\acmDOI{10.1145/3832783.3834593}
\acmYear{2026}
\copyrightyear{2026}
\acmISBN{979-8-4007-2882-2/2026/10}
\acmConference[ASE '26]{Proceedings of the 41st IEEE/ACM International Conference on Automated Software Engineering}{October 12--16, 2026}{Munich, Germany}
\acmBooktitle{Proceedings of the 41st IEEE/ACM International Conference on Automated Software Engineering (ASE '26), October 12--16, 2026, Munich, Germany}
\acmSubmissionID{ase26tool-p2-p}
\received{2026-05-11}
\received[accepted]{2026-06-19}

\usepackage{comment}
\usepackage{microtype}
\usepackage{xcolor}
\usepackage{xspace}

\newcommand{\approach}{\textsc{Graphectory Viewer}\xspace}

\newcommand{\graph}{\textsc{Graphectory}\xspace}

\newif\ifcomments
\commentsfalse

\begin{document}

\title[Graphectory Viewer]{Graphectory Viewer: A Tool for Process-Centric Analysis of Agentic Software Trajectories}

\author{Charlie Jyu}
\correspondingauthor
\orcid{0009-0007-1455-5129}
\affiliation{%
  \institution{University of Illinois at Urbana-Champaign}
  \city{Champaign}
  \country{USA}
}
\email{cbjyu2@illinois.edu}

\author{Shuyang Liu}
\orcid{0009-0009-7264-268X}
\affiliation{%
  \institution{University of Illinois at Urbana-Champaign}
  \city{Champaign}
  \country{USA}
}
\email{sl225@illinois.edu}

\author{Reyhaneh Jabbarvand}
\orcid{0000-0002-0668-8526}
\affiliation{%
  \institution{University of Illinois at Urbana-Champaign}
  \city{Champaign}
  \country{USA}
}
\email{reyhaneh@illinois.edu}

\begin{abstract}
We present \approach, a web-based tool for interactive, process-centric analysis of software-agent trajectories. Building on the Graphectory representation introduced in our previous work, \approach transforms heterogeneous raw trajectories into phase-aware graphs that connect low-level execution details with higher-level behavioral structures. The tool supports trajectories from multiple agent frameworks and provides interactive graph construction; node-level inspection of thoughts, actions, and observations; search and filtering over large trajectory collections; and Sankey-style summaries of problem-solving phase transitions. These capabilities enable researchers and practitioners to inspect individual executions, identify recurring behavioral patterns, compare successful and failed runs, and analyze large trajectory corpora beyond final task outcomes. To support reproducibility and further research, we release \approach as an open-source artifact together with documentation, precomputed graphs, and the large-scale trajectory corpus.

\noindent\textbf{Code:} \url{https://github.com/Intelligent-CAT-Lab/Graphectory}
\newline
\textbf{Dataset:} \url{https://doi.org/10.5281/zenodo.17364210}
\newline
\textbf{Live Demo:} \url{https://graphectory-viewer-demo.vercel.app/}
\newline
\textbf{Screencast:} \url{https://youtu.be/Hc4hnfRkuxc}
\end{abstract}


\maketitle

\section{Introduction}
\begin{figure*}[t]
    \centering
    \includegraphics[width=0.99\textwidth]{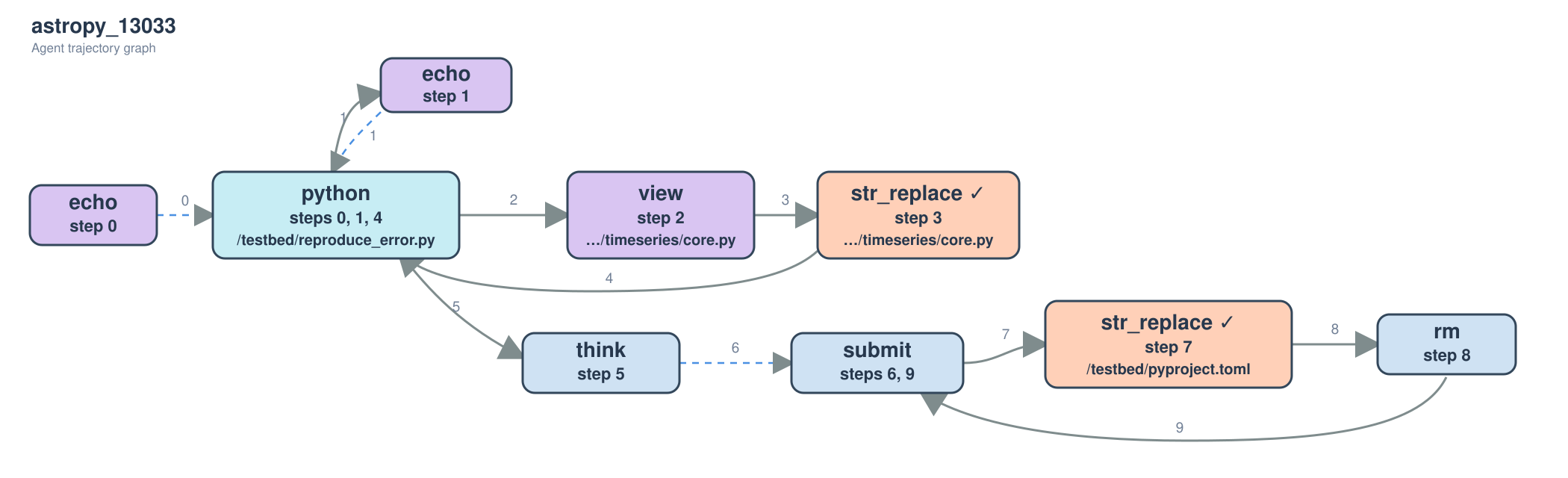}
    \Description{A phase-colored Graphectory graph for an agent trajectory repairing astropy issue 13033, with nodes for commands and thoughts connected by solid execution edges and dashed intra-step links.}
    \caption{The Graphectory for a SWE-agent trajectory repairing \texttt{astropy-13033}. Phase-colored nodes summarize actions, solid edges show execution flow, and dashed blue edges indicate intra-step command links.}
    \label{fig:astropy13033}
\end{figure*}

\label{sec:introduction}
Software engineering agents increasingly solve real-world programming tasks, including bug localization, code editing, and test generation. Tools such as SWE-agent\cite{yang2024sweagent}, OpenHands\cite{openhands2024}, and mini-SWE-agent\cite{minisweagent_docs} demonstrate autonomous issue resolution across complex code bases with limited human intervention. However, as these systems become more capable, their executions become harder to understand.

Most existing inspection tools present trajectories as raw logs or step-by-step transcripts, which are useful for replaying individual runs but provide limited support for higher-level analysis. In practice, researchers and practitioners need to answer questions such as: Did the agent correctly localize the bug before editing? Did it repeatedly explore the same files? Did it effectively validate its patch? How do the strategies of successful and failed runs differ? Answering these questions requires structured views that connect individual actions to higher-level problem-solving phases and support comparison across executions.

Prior work introduced \graph \citep{liu2025process}, a process-centric representation that maps low-level agent actions into semantic phases and graph structures, enabling analyses beyond outcome-centric evaluation. While effective for large-scale, automated studies, it is primarily used as an offline framework. Applying it requires custom scripts and provides limited support for interactively exploring graph structures.

We introduce \approach, an interactive, browser-based tool that operationalizes process-centric trajectory analysis by linking low-level execution evidence to higher-level behavioral structures. Given a raw trajectory, \approach normalizes framework-specific actions, assigns context-sensitive phase labels, and constructs an interactive, phase-aware directed graph. Users can inspect the thoughts, actions, and observations associated with each node, navigate repeated action occurrences, and suppress low-signal commands to reduce visual clutter. At the corpus level, a companion Sankey diagram aggregates phase transitions across trajectories, supporting the comparison of problem-solving strategies between runs, models, and outcomes. The tool also provides precomputed graphs and trajectories from the 4,000 attempted runs studied in~\cite{liu2025process}, of which 3,973 produced nonempty trajectories, facilitating replication and further analysis.

In summary, this paper makes the following contributions:
\begin{itemize}
    \item \textbf{Interactive process-centric trajectory analysis}
    (\S\ref{sec:overview}): We introduce \approach, a browser-based tool that connects low-level actions with higher-level behavioral structures through interactive, phase-aware graphs, and aggregate phase-transition summaries.

    \item \textbf{Preliminary user evaluation}
    (\S\ref{sec:user_study}): We conduct a within-subject study with five participants comparing \approach against the SWE-agent command-line viewer on five trajectory-forensics tasks. \approach increased aggregate task accuracy from 8\% to 84\%, and participants consistently preferred its visual workflow.

    \item \textbf{An open and reusable analysis artifact}
    (\S\ref{sec:utility}): We release the implementation, graph-generation pipeline, documentation, precomputed graph artifacts, and a large-scale trajectory corpus to support reproducible analysis and further research on software-agent behavior.
\end{itemize}

\section{Related Work}
\label{sec:related}

\paragraph{Software Agents.}
The rapid development of agentic software engineering has motivated growing interest in understanding and evaluating agent behavior beyond task-level success~\cite{hou2024large, abou2025agentic, wang2025ai, jiang2025agentic, yehudai2025survey, staufer2026agentindex}. Recent evaluation frameworks examine complementary aspects of agent behavior, including instruction following, alignment with intended goals, and reward design~\cite{gunnu2025cife, raghavendra2026agentic, akshathala2025beyond, zhu2025agenticbenchmark, koch2026beyondtask}. While these approaches provide metrics and protocols for evaluating agent capabilities, \approach supports interactive, process-centric analysis of how software agents navigate, modify, and validate code throughout their execution.

\paragraph{Agent Trajectory Visualization.}
Existing tools support trajectory inspection and visualization for different analysis and debugging needs. Framework-specific interfaces, such as the SWE-agent Trajectory Inspector and mini-SWE-agent viewer, and tools such as SeaView primarily support replay and inspection of individual executions as formatted text~\cite{sweagent_inspector_docs, minisweagent_docs, bula2025seaview}. AgentDiagnose provides general-purpose trajectory diagnostics through execution traces and embedding-based visualizations~\cite{ou2025agentdiagnose}, while AgentLens and AGDebugger support visual analysis, debugging, and steering of LLM-based agent systems~\cite{lu2024agentlens, epperson2025interactive}. In contrast, \approach maps heterogeneous software-agent trajectories into a graphical process-centric representation. It also supports corpus-level comparison through aggregate phase-transition summaries.

\begin{figure*}[t]
    \centering
    \includegraphics[width=0.86\textwidth]{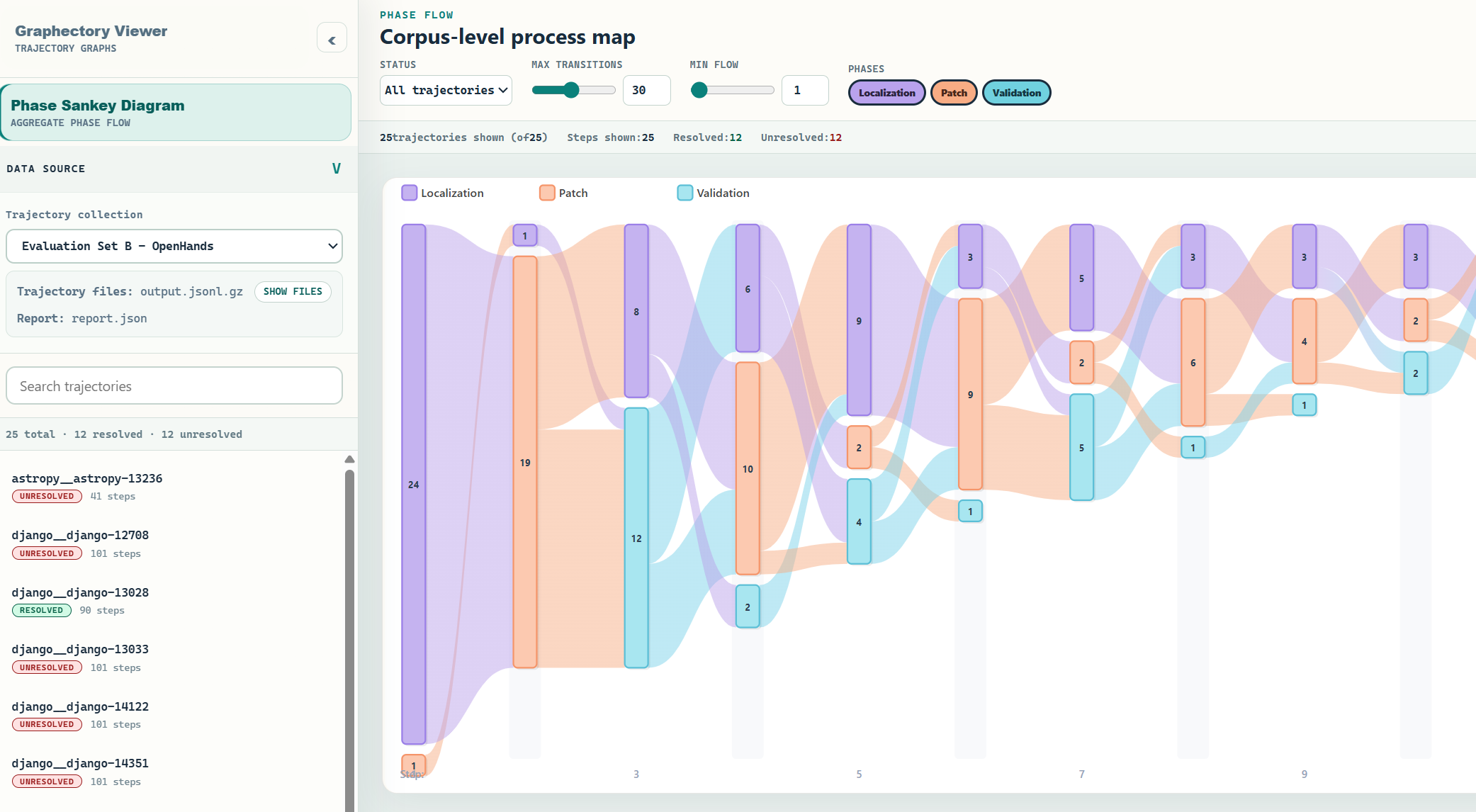}
    \Description{Screenshot of the Graphectory Viewer Phase Sankey Diagram interface, showing phase transitions, filters, and a searchable trajectory list.}
    \caption{The Phase Sankey Diagram interface with status filtering, phase chips, and sliders for transition depth and minimum displayed flow.}
    \label{fig:sankeyscreenshot}
\end{figure*}

\section{Tool Overview}
\label{sec:overview}
The main browser interface is organized around a navigation sidebar and a main visualization pane. As shown in Figure~\ref{fig:sankeyscreenshot}, users begin with the \textit{Data source} panel by supplying the trajectories and report. The sidebar then populates a searchable instance list with status badges and trajectory counts.

For trajectory-level analysis, selecting an instance from the sidebar loads a Graphectory into the main canvas. Figure~\ref{fig:astropy13033} shows a graphectory of SWE-agent using DeepSeek to solve issue astropy-13033 from SWE-bench Verified. The agent begins (step 0) by creating a script to replicate the issue from the problem description and running it with python. Steps 0-2 across the first four purple nodes represent the localization steps the agent takes to find the source of the issue. The second node (representing steps 0,1,4) is colored both as purple for localization and blue for validation because the command it represents occurs across multiple trajectory steps that belong to different phases. Step 3 shows the agent patching the code after localizing the issue then validating its correctness in step 4 by running the issue replication script. After thinking in step 5 about step 4's validation results, the model submits, briefly cleaning up the code in steps 7 and 8 before finally submitting. Solid edges show the main execution flow, while dashed blue edges represent intra-step links between multiple commands issued within the same trajectory step. Arrowhead size of an edge shows the length of the thought in the node it points to.

Clicking a node opens the inspection sidebar on the right and the file footprint menu. The sidebar shows the raw step's thought-action-observation tuple. When a node is referenced by multiple steps, the interface provides tabs for each step. The file footprint appears on the left of the canvas to show a visual representation of the files viewed and edited over the course of the trajectory. It also shows when each file was interacted with, providing a clear file level representation of how the agent acted.

The Phase Sankey Diagram, shown in Figure~\ref{fig:sankeyscreenshot}, provides the second major interface mode by summarizing each run as a sequence of meaningful phase transitions. Users can filter the displayed trajectories by status, adjust the maximum number of displayed transitions, suppress low-frequency flows with a minimum-flow slider, and enable or disable entire phase categories.

The backend converts trajectory directories or OpenHands-style \texttt{output.jsonl} files into a shared process-centric graph representation. First, \texttt{commandParser.py} uses Bash-aware parsing to normalize shell commands, chained actions, and framework-specific tool calls into structured records~\cite{bashlex}. Next, \texttt{mapPhase.py} assigns context-sensitive localization, patch, validation, or general labels following the Graphectory taxonomy~\cite{liu2025process}; for example, \texttt{pytest} may represent localization before an edit and validation afterward. Finally, \texttt{buildGraph.py} deduplicates repeated actions into shared nodes while retaining their information in the graph. The same pipeline supports serialized graph export, while the Sankey visualizer collapses repeated neighboring phases into compact paths for corpus-level comparison.

\section{User Evaluation}
\label{sec:user_study}

We conducted a preliminary within-subject usability study with five participants whose backgrounds ranged from basic computer science knowledge to professional software development and AI experience. Participants were recruited through convenience sampling and were told that the interfaces, rather than their individual ability, were being evaluated. Less experienced participants received a brief introduction to software agents, SWE-agent trajectories, and the fields shown by the viewers.

Each participant used both the SWE-agent command-line trajectory viewer and Graphectory Viewer to complete the same five trajectory-forensics tasks. The tasks required identifying a trajectory containing: (1) a command repeated at least 20\,times, (2) thought continuation across consecutive steps, (3) at least two command loops, each repeating the same command and parameters at least three times, (4) a non-submission thought longer than 700\,characters, and (5) an observation longer than 20,000\,characters.

We created two evaluation sets of 25 trajectories generated using SWE-agent with Mistral Devstral Small on SWE-bench Verified tasks. For each participant, one set was assigned to the command-line viewer and the other to Graphectory. We alternated the assignment across participants to reduce bias from differences between the datasets. Participants used the command-line viewer first and Graphectory second, with a 10\,minute limit for all five tasks in each condition. We recorded the number of correct answers and the completion time when all tasks were finished before the limit. Participants were then invited to rate both interfaces and provide open-ended feedback.

The command-line viewer yielded 2 correct answers across 25 attempts (8\%), and no participant completed all five tasks within 10\,minutes. Graphectory yielded 21 correct answers (84\%); three participants completed all five tasks in 5:13.80, 7:10, and 7:58. Participants rated Graphectory between 8 and 9 out of 10 and consistently preferred its visual presentation. They highlighted the usefulness of phase colors, node and arrow encodings, raw-step inspection, and the Sankey view. Suggested improvements included a clearer legend, more accurate arrow scaling, and a tutorial; all of which we subsequently added.

These results provide preliminary evidence that Graphectory improves performance on the studied trajectory-forensics tasks. However, the small convenience sample, fixed interface order, and task design specifically targeting queries that may favor visual search limit generalizability of these results.

\section{Utility and Quality}
\label{sec:utility}

% \paragraph{Evidence of utility.}
To evaluate the tool more systematically, we measured it over the released trajectory corpus used in our broader Graphectory workflow. This corpus contains 3,973 trajectories spanning eight collections (27 of 4,000 runs resulted in empty trajectories): four SWE-agent runs and four OpenHands runs over SWE-bench Verified tasks. Across the full corpus, raw trajectories average 35,664\,lines and 3,047,249\,characters, but the corresponding Graphectories average only 34.53\,nodes and 49.60\,edges. Even at this scale, the viewer preserves the underlying evidence because each node remains linked to its full thought, action, and observation tuples.

Table~\ref{tab:utility-stats} shows that this compression is consistent across agents and models, while also revealing meaningful behavioral differences. For example, SWE-agent with DeepSeek-V3 yields especially compact graphs, averaging only 14.99\,nodes and a shortest execution path of 7.50 despite raw trajectories averaging 15,895\,lines. At the other extreme, SWE-agent with Claude Sonnet 4 produces much larger logs, averaging 144,587\,lines and over 10\,million\,characters, yet the corresponding graphs still average only 46.49\,nodes. Deduplication is substantial across all collections, reducing effective graph size by 9.60\% to 28.22\% on average depending on the source. These reductions quantify how repeated actions are merged into shared nodes, revealing the backtracking and looping behavior that would otherwise be hidden in long, linear trajectory logs.

Beyond compression, the viewer enabled several recurring behavioral patterns to become apparent across large trajectory collections. We frequently observed agents entering repeated localization or validation loops on difficult tasks, repeatedly revisiting the same actions without transitioning into new phases or making meaningful progress. The process-centric representation additionally revealed that thought length alone is an unreliable proxy for reasoning complexity: agents sometimes execute complex multi-stage commands following only shallow thoughts, while lengthy reasoning traces may precede relatively trivial actions. These patterns were difficult to identify consistently from raw transcripts alone, but become substantially more visible through aggregate graph and phase-transition analysis.

\begin{table}[t]
\centering
\caption{Trajectory and graph average statistics across released SWE-agent and OpenHands collections.}
\label{tab:utility-stats}
\begin{tabular}{lrrrrr}
\toprule
Source & $N$ & lines & nodes & dedup. & path \\
\midrule
All collections & 3973 & 35,664 & 34.53 & 14.71\% & 16.70 \\
OH Claude-4     & 500  & 2,004  & 59.90 &  9.60\% & 30.49 \\
OH DeepSeek-V3  & 500  & 5,219  & 25.98 & 11.87\% & 13.75 \\
OH DeepSeek-R1  & 474  & 3,681  & 19.46 & 11.07\% & 12.57 \\
OH Devstral     & 500  & 12,446 & 72.53 & 10.37\% & 35.66 \\
SA DeepSeek-V3  & 499  & 15,895 & 14.99 & 20.71\% &  7.50 \\
SA Claude-4     & 500  & 144,587 & 46.49 & 11.94\% & 16.35 \\
SA DeepSeek-R1  & 500  & 7,466  & 12.57 & 13.71\% &  6.40 \\
SA Devstral     & 500  & 92,314 & 23.50 & 28.22\% & 10.66 \\
\bottomrule
\end{tabular}
\end{table}

\section{Conclusion}
\label{sec:Conclusion}

We presented \approach, an interactive tool that transforms heterogeneous software-agent trajectories into phase-aware graphs and aggregate transition summaries. Our evaluation provides preliminary evidence that the viewer can support process diagnosis over raw-log inspection in the studied tasks, while the corpus analysis illustrates how its phase-aware representation can expose agent problem-solving strategies. While the current system relies on domain-specific mapping rules, its adapter-based architecture facilitates extension to new domains and agent frameworks. By releasing the implementation, documentation, and precomputed corpus, we aim to support reproducible analysis and debugging of increasingly complex software-agent executions.

\section{Data Availability Statement}
\label{sec:data_availability}

The code and viewer are publicly available under the University of Illinois/NCSA Open Source License~\cite{graphectory-viewer-code}. The raw trajectory corpus is archived in Zenodo~\cite{graphectory-dataset} under DOI \href{https://doi.org/10.5281/zenodo.17364210}{10.5281/zenodo.17364210}. The repository provides installation instructions, usage examples, sample inputs, graph-export scripts, trajectory-metric scripts, and a screencast. A live demo is available at \url{https://graphectory-viewer-demo.vercel.app/}. 
%The artifact supports extension to new agent tools and new trajectory schemas, making it reusable as agent frameworks, model backbones, and orchestration strategies evolve.

\bibliographystyle{ACM-Reference-Format}
\bibliography{references}

\end{document}